\documentclass[a4paper]{article}
\usepackage{amsmath,graphicx,amssymb,delarray,subfigure,verbatim, bbding,mathrsfs}
\usepackage[colorlinks,linkcolor=black]{hyperref}
\usepackage{INTERSPEECH2019}

\title{A Recursive Network with Dynamic Attention for Monaural Speech Enhancement}

\name{Andong Li$^{1,3}$, Chengshi Zheng$^{1,3}$, Cunhang Fan$^{2, 3}$, Renhua Peng$^{1,3}$, Xiaodong Li$^{1,3}$}

\address{
	$^1$ Key Laboratory of Noise and Vibration Research, Institute of Acoustics, Chinese Academy\\
	of Sciences, Beijing, China\\
	$^2$ NLPR, Institute of Automation, Chinese Academy of Sciences, Beijing, China\\
	$^3$University of Chinese Academy of Sciences, Beijing, China}
\email{\{liandong, cszheng, pengrenhua, lxd\}@mail.ioa.ac.cn, cunhang.fan@nlpr.ia.ac.cn}

\begin{document}

\maketitle
\begin{abstract}
For continuous speech processing, dynamic attention is helpful in preferential processing, which has already been shown by the auditory dynamic attending theory. Accordingly, we propose a framework combining dynamic attention and recursive learning together for monaural speech enhancement. Apart from a major noise reduction network, we design a separated sub-network, which adaptively generates the attention distribution to control the information flow throughout the major network. Recursive learning is introduced to dynamically reduce the number of trainable parameters by reusing a network for multiple stages, where the intermediate output in each stage is corrected with a memory mechanism.
By doing so, a more flexible and better estimation can be obtained. We conduct experiments on TIMIT corpus. Experimental results show that the proposed architecture obtains consistently better performance than recent state-of-the-art models in terms of both PESQ and STOI scores. The code is provided at \url{https://github.com/Andong-Li-speech/DARCN}.

\end{abstract}
\vspace{-0.2cm}
\section{Introduction}
\vspace{-0.1cm}
\setlength{\parskip}{0.2em}
In real environments, clean speech is often contaminated by background interference, which may significantly reduce the performance of automatic speech recognition~{\cite{graves2013speech}}, speaker verification~{\cite{reynolds2000speaker}} and hearing aids~{\cite{dillon2008hearing}}. Monaural speech enhancement aims to extract the target speech from the mixture when only one-microphone is available~{\cite{loizou2013speech}}. In recent years, deep neural networks (DNNs) have shown their promising performance on monaural speech enhancement even in highly non-stationary noise environments owing to their superior capability in modeling complex nonlinearity~{\cite{wang2018supervised}}. Typical DNN-based methods can be categorized into two classes according to estimation targets, where one is masking-based~{\cite{wang2014training}} and the other is spectral mapping-based~{\cite{xu2014regression}}.

Conventional DNNs usually adopted fully-connected (FC) layers for noise reduction~{\cite{wang2014training, xu2014regression}}. To tackle the speaker generalization problem, Chen \textit{et al.} proposed to utilize stacked long short-term memory (SLSTM)~{\cite{chen2016large}}, which significantly outperformed DNNs. Recently, various convolutional neural networks (CNNs) with complex topology were proposed~{\cite{fu2017complex, tan2018gated, pandey2019tcnn, pirhosseinloo2019monaural}}, which could reduce the number of trainable parameters. More recently, Tan \textit{et al.} combined convolutional auto-encoder (CAE)~{\cite{badrinarayanan2015segnet}} and LSTM together to propose convolutional recurrent neural network (CRN)~{\cite{tan2018convolutional}}, where CAE helped to learn temporal-frequency (T-F) patterns and LSTM effectively captured dynamic sequence correlations.  

A variety of models with more complex topologies have been proposed recently~{\cite{tan2018gated, pandey2019tcnn, pirhosseinloo2019monaural, tan2018convolutional}}, which have shown improved performance, they still have some limitations for the following two fold. For one thing, the number of the parameters is often partly limited to meet the low-latency requirement, which heavily restricts the depth of the network. For another, the increase of depth is more likely to cause a gradient vanishing problem. Recently, progressive learning was proposed~{\cite{gao2016snr, li2019convolutional}}, which decomposes the mapping process into multiple stages. Experimental results in~{\cite{li2019convolutional}} have shown that it dramatically decreases the number of trainable parameters and effectively maintains the performance. Based on this conception, recursive learning~{\cite{li2020recursive}} was proposed by reusing the network for multiple stages, where the output in each stage is linked by a memory mechanism. It further alleviates the parameter burden and deepens the network without introducing extra parameters. 

Human tends to generate adaptive attention with dynamic neuron circuits to percept complicated environments~{\cite{anderson2013dynamic}}, which is also described by the auditory dynamic attending theory for continuous speech processing~{\cite{jones1976time}}. For example, when a person hears an utterance from the real environments, the more noise components are dominant, the more neuron attentions are needed to figure out the meaning and vice versa. This phenomenon reveals the dynamic mechanism of auditory perception system. Motivated by the physiological phenomenon, we propose a novel network combining dynamic attention and recursive learning together. Different from the previous networks~{\cite{tan2018gated, pandey2019tcnn, pirhosseinloo2019monaural, tan2018convolutional}} that a single complex network is designed for the task, the framework encompasses a major network and an auxiliary sub-network in parallel, where the one is noise reduction module (NRM) and the other is attention generator module (AGM). The workflow of the framework is as follows: at each intermediate stage, both the noisy feature and the estimation from the last stage are combined into the current input. AGM is adopted to generate the attention set, which is subsequently applied to NRM through the pointwise convolution and sigmoid function. In this way, AGM actually serves as a type of perception module to flexibly adjust the weight distribution of NRM, leading to better performance for noise suppression. To our best knowledge, it is the first time for a dynamic attention mechanism to be introduced for speech enhancement task.

The remainder of the paper is structured as follows. Section~{\ref{problem-formulation}} formulates the problem. The architecture of the network is illustrated in Section~{\ref{architecture-illustration}}. Section~{\ref{experimental-setup}} is the dataset and experimental settings. Section~{\ref{results-and-analysis}} presents the results and analysis. Section~{\ref{coclusions}} draws some conclusions.

\section{Problem formulation and notation}
\vspace{-0.2cm}
\label{problem-formulation}
In the time domain, a noisy signal can be modeled as $x \left( n \right) = s \left( n \right) + d \left( n \right)$, where $n$ is the discrete time index. With short-time Fourier transform (STFT), it can be further rewritten as:
\begin{equation}
\label{eqn:equa1}
X_{k, \ell} = S_{k, \ell} + D_{k, \ell},
\end{equation}
where $X_{k, \ell}$, $S_{k, \ell}$, and $D_{k, \ell}$ respectively refer to the noisy, clean, and noise components at the frequency bin index $k$ and the time frame index $\ell$. In this study, the network is deployed to estimate the magniude of spectrum (MS), which is then incorporated with noisy phase to recover the estimated spectrum. Inverse short-time Fourier transform (iSTFT) is used to reconstruct the waveform in the time domain.

For simplicity of notation, we define the principal notations used in this paper. $\mathbf{| X |}\in\mathbb{R}^{T\times F}$, $\mathbf{| S |}\in\mathbb{R}^{T\times F}$, $| \mathbf{\tilde S}^{l} |\in\mathbb{R}^{T\times F}$, and $\mathbf{| \tilde S |}\in\mathbb{R}^{T\times F}$ denote the magnitude of noisy spectrum, the magnitude of clean spectrum, the estimated magnitude of spectrum in the $l$th stage, and the final estimated magnitude of clean spectrum, respectively. $T$ and $F$ refer to the timestep and the feature length, respectively. As recursive learning is used, the superscript $l$ denotes the stage index, and the number of stages is notated as $Q$.  

\vspace{-0.2cm}
\section{Architecture illustration}
\label{architecture-illustration}
\subsection{Stage recurrent neural network}
\vspace{-0.2cm}
Stage recurrent neural network (SRNN) is first proposed in~{\cite{li2020recursive}}, which is the core component in recursive learning. It is capable of aggregating the information across different stages with a memory mechanism, which is comprised of two parts, namely two-dimensional convolutional (2-D Conv) block and convolutional-RNN (Conv-RNN) block. The first part tries to project the input features into a latent representation, followed by Conv-RNN to update the state in the current stage. Assuming the output of 2-D Conv and Conv-RNN at stage $l$ are respectively notated as $\mathbf{\hat h}^{l}$ and $\mathbf{h}^{l}$, the inference of SRNN is formulated as:
\vspace{-0.1cm}
\begin{gather}
\label{eqn:equa2}
\mathbf{\hat h}^{l} = f_{conv} \left( \mathbf{|X|}, |\mathbf{\tilde{S}}^{l - 1}| \right),\\
\mathbf{h}^{l} = f_{conv\_rnn} \left( \mathbf{\hat h}^{l}, \mathbf{h}^{l-1} \right),
\end{gather}   
\vspace{-0cm}
where $f_{conv}$ and $f_{conv\_rnn}$ refer to the functions of 2-D Conv and Conv-RNN, respecively. $\mathbf{h}^{l-1}$ is the state of the last stage. In this study, ConvGRU~{\cite{ballas2015delving}} is adopted as the unit in Conv-RNN, whose calculation process gives as follows:
\setlength{\abovedisplayskip}{2pt}
\setlength{\belowdisplayskip}{2pt}
\begin{gather}
\label{eqn:equa3}
\mathbf{z}^{l} = \sigma\left(\mathbf{W}^{l}_{z}\circledast\mathbf{\hat h}^{l} + \mathbf{U}^{l}_{z}\circledast\mathbf{h}^{l-1} \right),\\
\mathbf{r}^{l} = \sigma\left(\mathbf{W}^{l}_{r}\circledast\mathbf{\hat h}^{l} + \mathbf{U}^{l}_{r}\circledast\mathbf{h}^{l-1} \right),\\
\mathbf{n}^{l} = \tanh\left(\mathbf{W}^{l}_{n}\circledast\mathbf{\hat h}^{l} + \mathbf{U}^{l}_{n}\circledast\left(\mathbf{r}^{l}\odot\mathbf{h}^{l-1}\right) \right),\\
\mathbf{h}^{l} = \left(\mathbf{1} - \mathbf{z}^{l}\right)\odot\mathbf{\hat h}^{l} + \mathbf{z}^{l}\odot\mathbf{n}^{l},
\end{gather}
where $\mathbf{W}$ and $\mathbf{U}$ represent the weight matrices of the cell. $\sigma(\cdot)$ and $\tanh(\cdot)$, respectively, denote the sigmoid and the tanh activation functions. $\circledast$ represents the convolutional operator and $\odot$ is the element-wise multiplication. Note that biases are ignored for notation convenience.
\vspace{-0.1cm}
\subsection{Attention gate}
\begin{figure}[t]
	\centering
	\centerline{\includegraphics[width=\columnwidth]{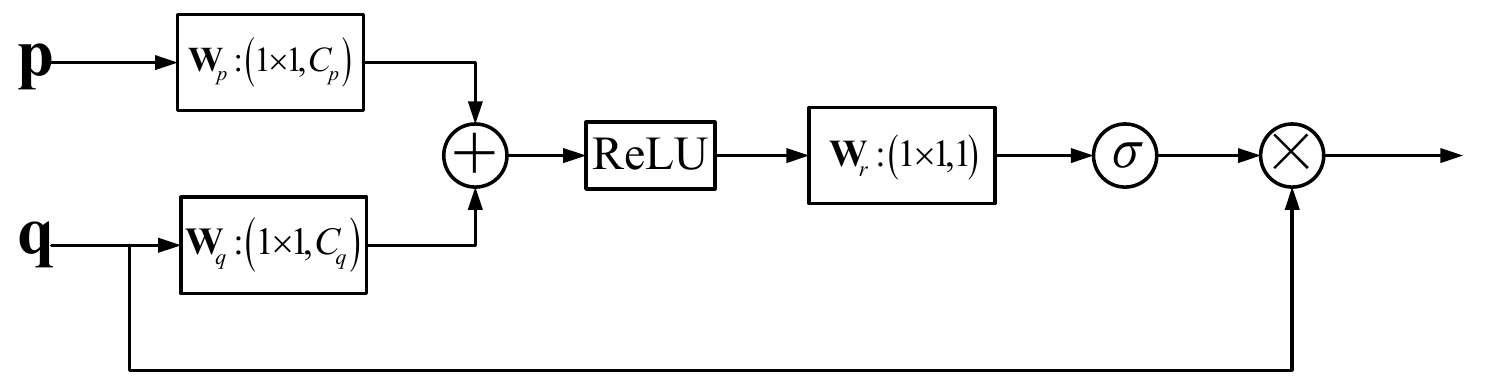}}
	\caption{The structure of attention gate adopted in NRM. $\mathbf{p}$ and $\mathbf{q}$ refer to the feature of a decoding layer and its corresponding feature in a encoding layer, respectively. $\mathbf{W_{p}}$, $\mathbf{W_{q}}$ and $\mathbf{W_{r}}$ are 2-D convolutional layers, whose kernel size are $1\times1$. $C_{p}$ and $C_{q}$ refer to the channel numbers of $\mathbf{p}$ and $\mathbf{q}$. Batch normalization is used after each convolutional operation.}
	\label{fig:attention-gate}
	\vspace{-0.6cm}
\end{figure}
\vspace{-0.2cm}
Attention U-Net (AU-Net) is first proposed in~{\cite{oktay2018attention}} to improve the accuracy in the segmentation-related tasks, where attention gates (AGs) are inserted between the convolutional encoder and the decoder. Compared with a standard U-Net, AU-Net has the capability of automatically suppressing the irrelevant regions and emphasizing the important features. As the spectrum includes abundant frequency components, where formants are often dominant in the low-frequency regions and the regions of the high-frequency have a sparse distribution, it is necessary to discriminate different spectral regions with different weights. The schematic of the AG adopted in this paper is shown in Figure~{\ref{fig:attention-gate}}. Assuming the inputs of the unit are $\mathbf{p}$ and $\mathbf{q}$, where $\mathbf{p}$ and $\mathbf{q}$ refer to the feature of a decoding layer and its corresponding feature in an encoding layer, respectively. the output can be calculated as:
\begin{equation}
\label{equa4}
\mathbf{y} = \mathbf{q}\odot \sigma \left( \mathbf{W}_{r}\circledast ReLU \left( \mathbf{W}_{p}\circledast \mathbf{p} + \mathbf{W}_{q}\circledast \mathbf{q} \right) \right),
\end{equation}
where $\mathbf{W}_{p}$, $\mathbf{W}_{q}$ and $\mathbf{W}_{r}$ are the convolution kernels. Note that the unit consists of two branches, where the one merges the information of both inputs and generates the attention coefficients through a sigmoid function and the other copy the information of $\mathbf{q}$ and multiply the coefficients. After the output of AG is obtained, it is concatenated with the feature from the corresponding decoding layer along the channel dimension as the input of the next decoding layer.

\begin{figure*}[t]
	\vspace{-0.3cm}
	\setlength{\abovecaptionskip}{0.235cm}
	\setlength{\belowcaptionskip}{-0.1cm}
	\centering
	\centerline{\includegraphics[width=145mm]{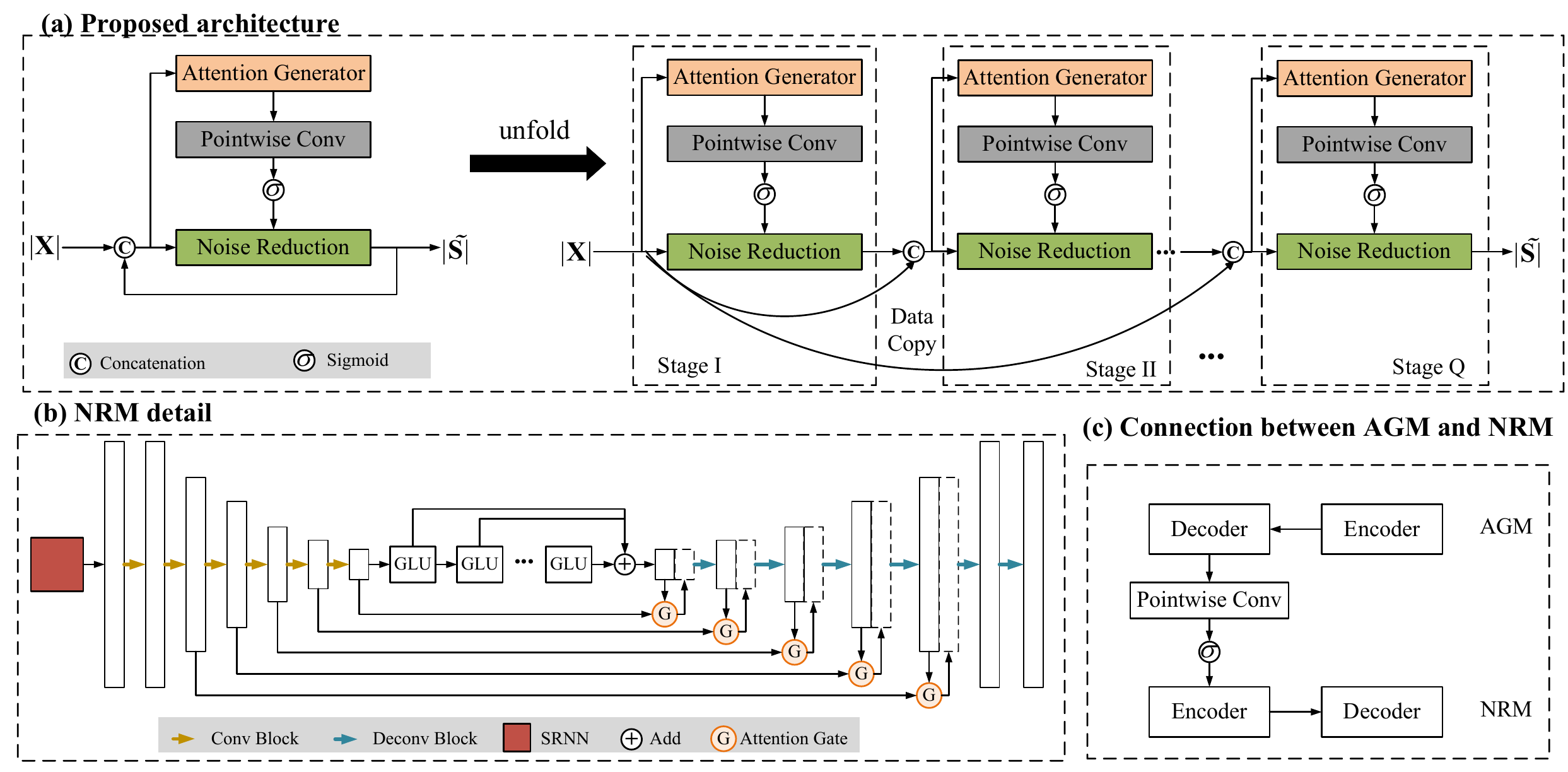}}
	\caption{The schematic of proposed architecture. (a) Proposed architecture and its unfolding structure. The architecture encompasses two parts in parallel, namely AGM and NRM. (b) The detailed structure of NRM. (c) The connection between AGM and NRM.}
	\label{fig:proposed-architecture}
	\vspace{-0.5cm}
\end{figure*}
\vspace{-0.2cm}
\subsection{Proposed architecture}
\vspace{-0.2cm}
The overview of the proposed architecture is depicted in Figure~{\ref{fig:proposed-architecture}}-(a). It has two modules, namely AGM and NRM, and the two modules are designed to interleave execution during the whole process. The architecture is operated with a recursive procedure, i.e., the whole forward stream can be unfolded into multiple stages. In each stage, the original noisy spectrum and the estimation from the last stage are concatenated, serving as the network input. It is sent to AGM to generate the current attention set, representing the attention distribution at the current stage. It is subsequently applied to NRM to control the information flow throughout the network. NRM also receives the input to estimate the MS. As a consequence, the output of AGM dynamically depends on how well MS is estimated in the last stage, i.e., AGM is capable of re-weighting the attention distribution according to the previous feedback from the noise reduction system. 

Assuming the mapping functions of AGM and NRM are denoted as $G_{A}$ and $G_{R}$, respectively. The calculation procedure of the proposed architecture works as follows:
\setlength{\abovedisplayskip}{1pt}
\setlength{\belowdisplayskip}{1pt}
\begin{gather}
\label{eqn:equa5}
\mathbf{a}^{l} = G_{A} \left( \mathbf{|X|}, |\mathbf{\tilde{S}}^{l-1}|;\theta_{A} \right),\\
|\mathbf{\tilde{S}}^{l}| = G_{R} \left( \mathbf{|X|}, |\mathbf{\tilde{S}}^{l-1}|, \mathbf{a}^{l} ;\theta_{R} \right),
\end{gather} 
where $\mathbf{a}^{l}$ is the generated attention set at stage $l$. $\theta_{A}$ and $\theta_{R}$ represent the network parameters for AGM and NRM. 

In this study, we use a typical U-Net~{\cite{ronneberger2015u}} topology for AGM, which consists of the convolutional encoder and the decoder. The encoder consists of five successive 2-D convolutional layers, each of which is followed by batch normalization (BN)~{\cite{ioffe2015batch}} and exponential linear unit (ELU)~{\cite{clevert2015fast}}. The number of channels through the encoder is (16, 32, 32, 64, 64). The decoder is the mirror representation of the encoder except all the convolutions are replaced by deconvolutions~{\cite{noh2015learning}} to effectively enlarge the mapping size. Similarly, the number of channels through the decoder is (64, 64, 32, 32, 16). The kernel size, stride for both encoder and decoder are (2, 5) and (1, 2), respectively. Skip connections are introduced to compensate for information loss during the encoding process.

The detail of NRM is shown in Figure~{\ref{fig:proposed-architecture}-(b). It includes three parts, namely SRNN, AU-Net and a series of GLUs~{\cite{tan2018gated}}. Given the input of the network, whose size is $T\times F\times2$. 161 is the feature length, and 2 is the number of the input channel. The output size after SRNN and consecutive six convolutional blocks are $T\times4\times64$. It is subsequently reshaped into $T\times256$. six concatenated GLUs proposed by~{\cite{tan2018gated}} are set to explore the contextual correlations efficiently. The output of GLUs is reshaped back to $T\times4\times64$, which is subsequently sent to the decoder to expand the feature size and estimate the MS. The number of channels for the encoder and decoder in AU-Net are (16, 16, 32, 32, 64, 64) and (64, 32, 32, 16, 16, 1), respectively. The kernel size and stride in NRM are the same as the setup in AGM except for the last layer, which takes the pointwise convolution, followed by Softplus as the nonlinearity~{\cite{zheng2015improving}} to obtain the MS. Note that different from direct skip connections in a standard U-Net, the feature mappings from the encoder are multiplied with the gating coefficients from the AGs before they are concatenated with the decoding features, which help to weigh the feature importance in multiple encoding layers.

The connection between AGM and NRM is shown in Figure~{\ref{fig:proposed-architecture}-(c), where each of the middle features in the decoder of AGM is multiplied to the corresponding feature in the encoder of NRM through the pointwise convolution and the sigmoid function. Note that the sigmoid function is applied to range the value scale into $\left( 0, 1 \right)$. 
\vspace{-0.2cm}
\subsection{Loss function}
\vspace{-0.1cm}
As the network is trained for multiple stages, at each of which we obtain an intermediate estimation, the accumulated loss can be defined as $\mathcal{L} = \sum_{l=1}^{Q}\lambda_{l}\mathcal{D}^{l} \left( \mathbf{\tilde{S}}^{l}, \mathbf{S} \right)$, where $\lambda_{l}$ is the weighted coefficient for each stage, $\mathcal{D}^{l} \left( \cdot \right)$ is the loss function for the $l$th stage. We set $\lambda_{l}\equiv 1$, with $l =1,\cdots Q$ in this study, i.e., the same emphasis is given to each training stage.
\vspace{-0.3cm}
\renewcommand\arraystretch{0.75}
\begin{table*}[t]
	\tiny
	\caption{Experimental results under seen noise cases. \textbf{BOLD} indicates the best result for each case. The number of stage Q = 3 for proposed architecture.}
	\centering
	\resizebox{\textwidth}{!}{
		
		\begin{tabular}{cccccc|ccccc}
			\hline
			\multicolumn{1}{c}{Metric} & \multicolumn{5}{c|}{PESQ}  & \multicolumn{5}{c}{STOI (in \%)} \\
			\hline
			\multicolumn{1}{c}{SNR} & -5\rm{dB}  &0\rm{dB} &5\rm{dB} &10\rm{dB} &Avg.
			& -5\rm{dB}  &0\rm{dB} &5\rm{dB} &10\rm{dB} &Avg.\\
			\hline
			\multicolumn{1}{c}{Noisy}  &1.46 &1.76 &2.14 &2.55 &1.98 &62.44 &71.55 &81.81 &89.53 &76.33 \\
			SLSTM &2.35 &2.67 &2.98 &3.21 &2.80 &80.16 &86.64 &91.24 &94.18 &88.06\\
			
			CRN &2.42 &2.74 &3.06 &3.34 &2.89 &81.45 &88.09 &92.60 &95.57 &89.43\\
			
			GRN &2.47 &2.76 &3.04 &3.25 &2.88 &82.94 &88.63 &92.76 &95.43 &89.94\\
			
			DCN &2.41 &2.75 &3.06 &3.31 &2.88 &80.64 &87.40 &92.10 &95.03 &88.79\\
			
			Proposed (Q = 3) &\textbf{2.60} &\textbf{2.91} &\textbf{3.21} &\textbf{3.46} &\textbf{3.05} &\textbf{83.74} &\textbf{89.07} &\textbf{93.17} &\textbf{95.79} &\textbf{90.44}\\
			\hline
	\end{tabular}}
	\label{tbl:seen-results}
	\vspace{-0.2cm}
\end{table*}

\renewcommand\arraystretch{0.75}
\begin{table*}[t]
	\caption{Experimental results under unseen noise cases. \textbf{BOLD} indicates the best result for each case. The number of stage Q = 3 for proposed architecture.}
	\centering
	\tiny
	\resizebox{\textwidth}{!}{
		
		\begin{tabular}{cccccc|ccccc}
			\hline
			\multicolumn{1}{c}{Metric} & \multicolumn{5}{c|}{PESQ}  & \multicolumn{5}{c}{STOI (in \%)} \\
			\hline
			\multicolumn{1}{c}{SNR} & -5\rm{dB}  &0\rm{dB} &5\rm{dB} &10\rm{dB} &Avg.
			& -5\rm{dB}  &0\rm{dB} &5\rm{dB} &10\rm{dB} &Avg.\\
			\hline
			\multicolumn{1}{c}{Noisy}  &1.47 &1.83 &2.13 &2.51 &1.98 &60.15 &72.51 &81.23 &89.77 &75.92 \\
			
			SLSTM &2.07 &2.50 &2.82 &3.12 &2.63 &74.23 &84.52 &89.69 &93.51 &85.49\\
			
			CRN &2.20 &2.60 &2.92 &3.21 &2.73 &77.10 &86.19 &91.03 &94.79 &87.28\\
			
			GRN &2.29 &2.63 &2.89 &3.16 &2.74 &78.10 &86.92 &91.28 &94.78 &87.77\\
			
			DCN &2.20 &2.62 &2.94 &3.22 &2.74 &76.05 &85.88 &90.73 &94.48 &86.78\\
			
			Proposed (Q = 3) &\textbf{2.37} &\textbf{2.75} &\textbf{3.06} &\textbf{3.34} &\textbf{2.88} &\textbf{79.16} &\textbf{87.54} &\textbf{91.61} &\textbf{95.02} &\textbf{88.33}\\
			\hline
	\end{tabular}}
	\label{tbl:unseen-results}
	\vspace{-0.3cm}
\end{table*}
\vspace{-0cm}

%
%
%
%
%
%
%
%
%
%
%
%
%

\section{Experimental setup}
\label{experimental-setup}
\subsection{Dataset}
\vspace{-0.2cm}
The experiments are conducted on TIMIT corpus~{\cite{garofolo1993darpa}}. 4856, 800 and 100 clean utterances are selected for training, validation and testing, respectively. Training and validation dataset are created under the SNR levels ranging from -5\rm{dB} to 10\rm{dB} with the the interval 1\rm{dB} whilst we test the model under the SNR conditions of (-5\rm{dB}, 0\rm{dB}, 5\rm{dB}, 10\rm{dB}). 130 types of noises used in~{\cite{li2020recursive}} are for training and validation. Another 5 types of noises from NOISEX92 (babble, f16, factory2, m109 and white) are used to explore the generalization capacity of networks. All the collected noises are first concatenated into a long vector. During each mixed process, a random cutting point is generated, which is subsequently mixed with an utterance under an SNR level. As a result, we create 40,000, 4000, 800 noisy-clean pairs for training, validation and testing, respectively. 
\vspace{-0.2cm}
\subsection{Baselines}
\vspace{-0.2cm}
In this study, four networks are selected as the baselines, namely SLSTM~{\cite{chen2016large}}, CRN~{\cite{tan2018convolutional}}, GRN~{\cite{tan2018gated}} and DCN~{\cite{pirhosseinloo2019monaural}}, all of which have achieved state-of-the-art performance recently. For SLSTM, four LSTM layers with 1024 units are stacked, followed by one FC layer to obtain the MS. The input of SLSTM includes the concatenation of the current frame and the previous ten frames. CRN is a type of real-time architecture combining CNN and LSTM. GRN and DCN are typical fully-convolutional network with gating mechanism. 
\vspace{-0.2cm}
\subsection{Parameter setup}
\vspace{-0.2cm}
All the utterances are sampled at 16kHz. The 20ms Hamming window is applied, with 10ms overlap in adjacent frames. 320-point STFT is adopted, leading to a 161-D feature vector in each frame. All the models are trained with mean-square error (MSE) criterion, optimized by Adam~{\cite{kingma2014adam}}. The learning rate is initialized at 0.001, we halve the learning rate when consecutive 3 validation loss increment happens and the training is early-stopped when 10 validation loss increment happens. All the models are trained for 50 epochs. The minibatch is set to 4 at an utterance level. Within a minibatch, the utterance whose timestep is less than the longest one is padded with zero.   
\vspace{-0.2cm}
\section{Results and analysis}
\label{results-and-analysis}
This section evaluates the performance of different models with perceptual evaluation speech quality (PESQ)~{\cite{recommendation2001perceptual}} and short-time objetive intelligibility (STOI) scores~{\cite{taal2010short}}.
\vspace{-0.2cm}
\begin{figure}[t]
	\setlength{\abovecaptionskip}{0.235cm}
	\setlength{\belowcaptionskip}{-0.1cm}
	\centering
	\subfigure[]{
		\includegraphics[width= 38mm]{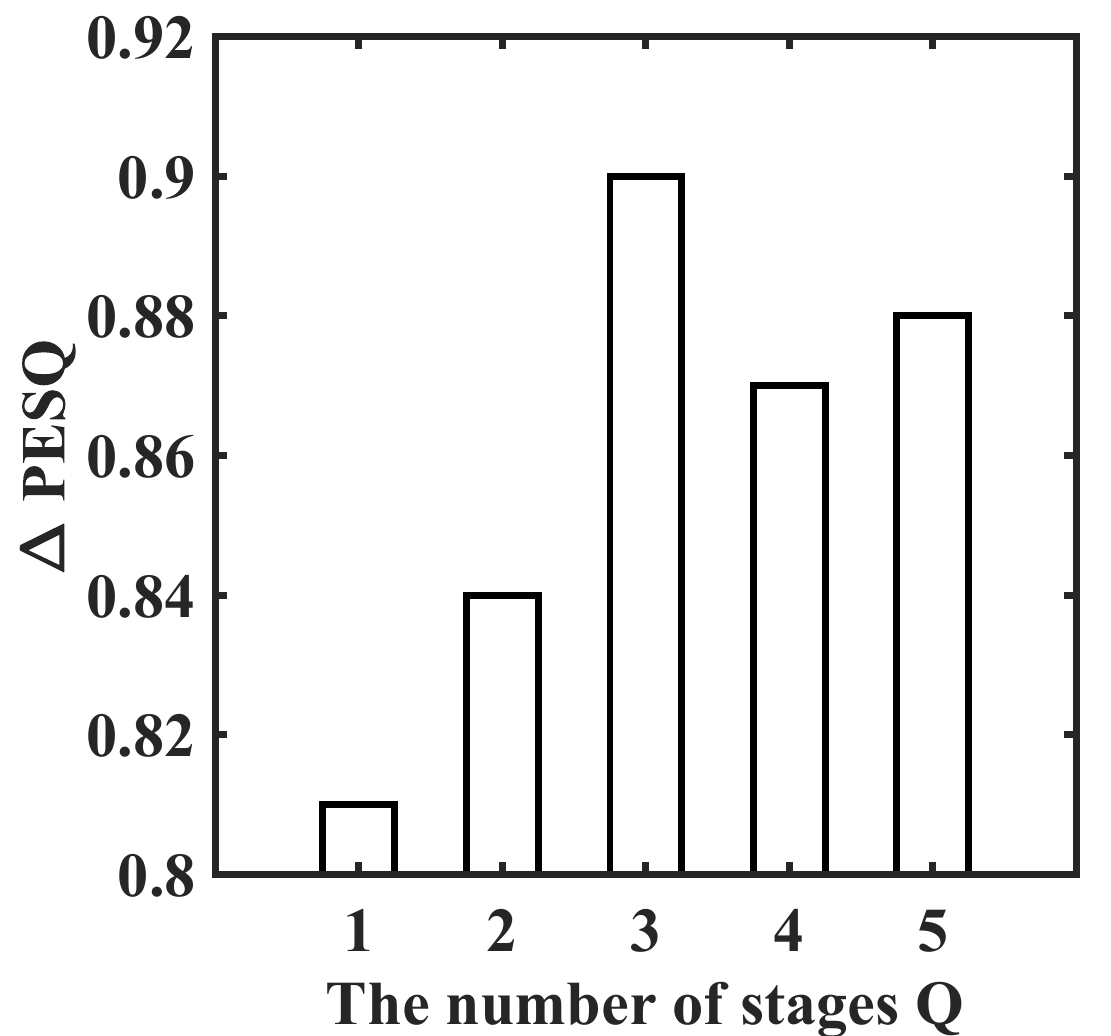}	
	}
	\subfigure[]{
		\includegraphics[width= 38mm]{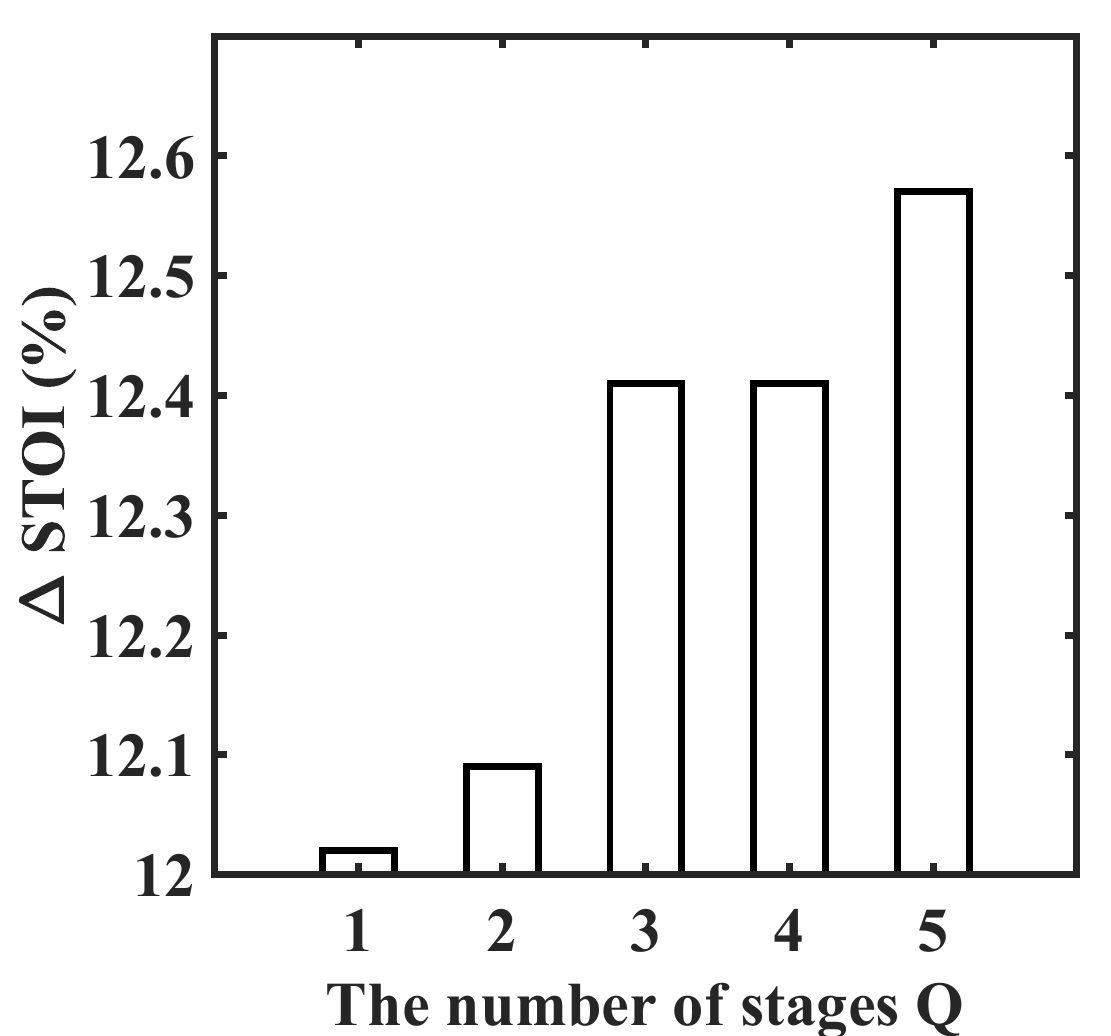}	
	}
	\caption{The impact of the number of stages Q. (a) PESQ improvement for different Q. (b) STOI improvement for different Q. All the values are evaluated for unseen noise cases and averaged over different SNRs.}
	\label{fig:result-stage}
	\vspace{-0.6cm}
\end{figure}

\subsection{Objective results}
\vspace{-0.2cm}
Tables~{\ref{tbl:seen-results}} and~{\ref{tbl:unseen-results}} summarize the results of different models for seen and unseen noise cases, respectively. From the two tables, one can observe the following phenomena. First, CRN, GRN, DCN, and the proposed model consistently outperform SLSTM in both seen and unseen noise cases. This is because SLSTM solely considers the sequence correlations but neglecting the implicit T-F patterns, which is crucial for spectrum recovery. Moreover, stacked LSTM tends to cause an attenuation effect due to the gradient vanishing problem, which limits the performance. Second, compared with the baselines, the proposed architecture obtains notable improvements in both metrics. For example, when going from CRN to the proposed model, PESQ is improved by 0.16 and STOI is improved by 1.01\% on average for seen cases. A similar trend is also observed for unseen cases, indicating that the proposed model has a good noise generalization capability. Third, we observe that GRN and DCN can achieve close performance. This can be explained as both networks have similar topology, where dilation convolutions combined with the gating mechanism are applied for sequence modeling.  
\vspace{-0.2cm}

\subsection{Impact of stages}
\vspace{-0.2cm}
We study the impact of the number of stages Q, which is given in 
Figure~{\ref{fig:result-stage}}. One can get that with the increase of Q, both values of PESQ and STOI are consistently improved when Q $\leq 3$. This indicates that SRNN can effectively refine the performance of the network by a memory mechanism. We also find that when Q increases from 3 to 5, PESQ value slightly degrades whilst STOI is still improved. This is because the distance-based loss like MSE is utilized, the loss function and the optimization process cannot guarantee consistent optimization for both metrics, which is consistent with the previous study in~{\cite{li2020recursive}}.  
\vspace{-0.4cm}

\renewcommand\arraystretch{1.0}
\begin{table}[t]
	\setlength{\abovecaptionskip}{0.235cm}
	\setlength{\belowcaptionskip}{-0.1cm}
	\caption{The number of trainable parameters among different models. The unit is million. $\textbf{BOLD}$ denotes the lowest trainable parameters. }
	\centering
	\footnotesize
	\begin{tabular}{|c|c|c|c|c|c|}
		\hline
		Model &SLSTM &CRN &GRN &DCN &Proposed\\
		\hline
		Para. (million) &36.81 &17.58 &3.13 &2.92 &\textbf{1.23}\\
		\hline
	\end{tabular}
	\label{tbl:model-parameters}
	\vspace*{-\baselineskip}
	\vspace{-0.6cm}
\end{table}

\subsection{Parameter comparison}
\vspace{-0.2cm}
Table~{\ref{tbl:model-parameters}} summarizes the number of trainable parameters for different models. One can see that the proposed model dramatically decreases the number of trainable parameters compared with other baselines. This demonstrates the superior parameter efficiency of the proposed architecture.

\vspace{-0.2cm}
\section{Conclusions}
\label{coclusions}
In a complicated scenario, a person usually dynamically adjusts the attention to the change of the environments for continuous speech. Based on this neural phenomenon, we propose a framework combining dynamic attention and recursive learning. To adaptively control the information flow of the noise reduction network, a separate sub-network is designed to update the attention representation in each stage and is subsequently applied to the major network. As a recursive paradigm is taken for training, the network is reused for multiple stages. As a result, we achieve a refined estimation stage by stage. Experimental results show that, compared with previous state-of-the-art strong models, the proposed model achieves consistently better performance while further decreasing the parameter burden.
   
\bibliographystyle{IEEEtran}

\bibliography{mybib}


\end{document}